\documentclass[traditabstract]{aa} % for the abstract without structuration 
\usepackage{gensymb}
\usepackage{amsmath}
\usepackage{graphicx}
\usepackage{txfonts}
\usepackage{natbib}
\usepackage{color}
\usepackage{multirow}
\usepackage{subfigure}
\usepackage{lscape}
\usepackage{latexsym}
\usepackage{changepage}         %for adjustwidth
\bibpunct{(}{)}{;}{a}{}{,} % to follow the A&A style
\usepackage[usenames,dvipsnames,svgnames,table]{xcolor}
\usepackage[breaklinks, colorlinks, citecolor=CornflowerBlue]{hyperref}
\usepackage{url}
\newcommand{\kms} {$\mbox{km s}^{-1}$}
\newcommand{\re}{$R_\mathrm{e}$}
\newcommand{\Msun} {$\mbox{M}_{\sun}$}

\newcommand{\Hb}{\hbox{H$\beta$}}
\newcommand{\Ha}{\hbox{H$\alpha$}}

\begin{document}

   \title{The ultra-diffuse galaxy NGC~1052-DF2 with MUSE: \\ I. Kinematics of the stellar body\thanks{Based on observations collected at the European Southern Observatory under ESO programmes 2101.B-5008(A) and 2101.B-5053(A).}} 
\titlerunning{NGC~1052-DF2 analysed with MUSE: I. Kinematics of the stellar body}

\author{Eric Emsellem\inst{1,2}, Remco F. J. van der Burg\inst{1}, J\'er\'emy Fensch\inst{1}, Tereza Je\v{r}\'abkov\'a\inst{1,3,4}, Anita Zanella\inst{1}, Adriano Agnello\inst{1,5}, Michael Hilker\inst{1}, Oliver Müller\inst{6}, Marina Rejkuba\inst{1}, Pierre-Alain Duc\inst{6}, Patrick Durrell\inst{7}, Rebecca Habas\inst{8}, Federico Lelli\inst{1}, Sungsoon Lim\inst{9}, Francine R. Marleau\inst{8}, Eric Peng\inst{10,11}, Rub\'en S\'anchez-Janssen\inst{12}
          }
   \authorrunning{Emsellem, van der Burg, Fensch, Je\v{r}\'abkov\'a, et al.}

   \institute{
   European Southern Observatory, Karl-Schwarzschild-Str. 2, D-85748 Garching, Germany
   \and
   Univ Lyon, Univ Lyon1, ENS de Lyon, CNRS, Centre de Recherche Astrophysique de Lyon UMR5574, F-69230 Saint-Genis-Laval France
      \and
   Helmholtz Institut f\"{u}r Strahlen und Kernphysik, Universit\"{a}t Bonn, Nussallee 14–16, 53115 Bonn, Germany
   \and
   Astronomical Institute, Charles University in Prague, V 
   Hole\v{s}ovi\v{c}k\'ach 2, CZ-180 00 Praha 8, Czech Republic
   \and
   DARK, Niels Bohr Institute, University of Copenhagen, Lyngbyvej 2, 2100 Copenhagen, Denmark
   \and
   Observatoire Astronomique de Strasbourg  (ObAS), Universit\'{e} de Strasbourg - CNRS, UMR 7550 Strasbourg, France
    \and
   Youngstown State University, One University Plaza, Youngstown, OH 44555 USA
      \and
   Institut f{\"u}r Astro- und Teilchenphysik, Universit{\"a}t Innsbruck, Technikerstra{\ss}e 25/8, Innsbruck, A-6020, Austria 
    \and
   NRC Herzberg Astronomy and Astrophysics Research Centre, 5071 West Saanich Road, Victoria, BC, V9E 2E7, Canada
   \and
   Department of Astronomy, Peking University, Beijing, China 100871
   \and
   Kavli Institute for Astronomy and Astrophysics, Peking University, Beijing, China 100871
   \and
   UK Astronomy Technology Centre, Royal Observatory, Blackford Hill, Edinburgh, EH9 3HJ, UK
             }

   \date{Received December 18, 2018}
% \abstract{}{}{}{}{} 
% 5 {} token are mandatory

   \abstract{
  
       The so-called ultra-diffuse galaxy NGC~1052-DF2 was announced to be a galaxy lacking dark matter based on a spectroscopic study of its constituent globular clusters. Here we present the first spectroscopic analysis of the stellar body of this galaxy using the MUSE integral-field spectrograph at the (ESO) Very Large Telescope. The MUSE datacube simultaneously provides DF2's stellar velocity field and systemic velocities for seven globular clusters (GCs). We further discovered three planetary nebulae (PNe) that are likely part of this galaxy. While five of the clusters had velocities measured in the literature, we were able to confirm the membership of two more candidates through precise radial velocity measurements, which increases the measured specific frequency of GCs in DF2. The mean velocity of the diffuse stellar body, 1792.9$^{-1.8}_{+1.4}$~\kms, is consistent with the mean globular cluster velocity. We detect a weak but significant velocity gradient within the stellar body, with a kinematic axis close to the photometric major axis, making it a prolate-like rotator. We estimate a velocity dispersion from the clusters and PNe of $\sigma_{\mathrm{int}}=10.6^{+3.9}_{-2.3}$~\kms. The velocity dispersion $\sigma_{\rm{DF2}\star}$(\re) for the stellar body within one effective radius is $10.8^{-4.0}_{+3.2}$~\kms. Considering various sources of systemic uncertainties, this central value varies between 5 and 13~\kms, and we conservatively report a 95\% confidence upper limit to the dispersion within one \re\ of 21~\kms. We provide updated mass estimates based on these dispersions corresponding to the different distances to NGC~1052-DF2 that have been reported in the recent literature.
 }
   \keywords{Galaxy: kinematics and dynamics, Galaxy: stellar content, Galaxies: dwarf
               }
   \maketitle
%
%________________________________________________________________

\hyphenation{in-tra-clus-ter}
\hyphenation{rank-or-der}

\section{Introduction}
\label{sec:intro}

Dwarf galaxies are by far the dominant galaxy population in numbers. They have been the target of extensive studies to probe the various formation channels that may be at work (\citealt{SandageBinggeli84, Kormendy85, Ferguson94, Moore98, Conselice03, Mastropietro05, Mayer2011, McConnachie2012, Skillman2012, Lelli2014}, and e.g. \citealt{Tolstoy2009} for a review). Within such a population, low-surface brightness (LSB) systems \citep{Bothun87, Impey88, Dalcanton95, Dalcanton97, Sprayberry95, Jimenez98, McGaugh96, Mateo98} have often been used as potential Rosetta Stones for our understanding of the cosmological paradigm  \citep[see e.g.][for a review]{Bullock17}.
Going from the low- to high-mass regimes of galaxies, it has been argued that the stellar mass build up mostly goes from in-situ to ex-situ \citep{Oser2010, Clauwens2018}. In other words, dwarfs tend to grow their stellar mass via internal star formation while galaxy mergers dominate the assembly of giant systems. 
Still, a common property for galaxies from low to high masses is the derived high dynamical mass-to-light ratios, something that is considered evidence for the presence of a significant fraction of dark matter. Another interesting emerging thread may be the existence at both extremes of a variety of morphological and dynamical structures, with for example kinematically decoupled components, or more generally departures from oblateness \citep[e.g.][]{Krajnovic2011, Battaglia2013, Toloba2014, Guerou2015}, which illustrates a probable richness of evolutionary paths. Scrutinising the internal stellar kinematics of galaxies has thus become an essential ingredient to constrain formation and evolution processes \citep[see e.g.][]{Cappellari2017}.

In such a context, two exceptional claims were made about NGC~1052-DF2 (DF2, hereafter), a so-called ultra-diffuse galaxy (UDG) likely located between 13~Mpc and 20~Mpc from us \citep{Trujillo2018,vanDokkum2018_dist}. First, by using the relative velocity distribution of ten spatially associated globular clusters (GC), \citet{vanDokkum2018_nat} inferred an internal velocity dispersion of less than 10~km/s for this galaxy and argued for a lack of dark matter in this galaxy assuming a distance of 20~Mpc. Second, based on GC numbers and mass estimates, the GC system of NGC~1052-DF2 would be over-populated and contain more massive clusters compared to what is expected for galaxies of similar halo mass \citep{vanDokkum2018_gc}. These claims triggered a strong response from the community, related to the statistical methods \citep{Martin2018}, the distance estimation \citep{Trujillo2018}, the reliability of estimated masses \citep{Laporte2018, Hayashi2018}, and the implications for modified gravity \citep{Famaey2018,kroupa2018}. With current uncertainties, the mass estimates range from a starlight-dominated to a dark-matter-dominated galaxy.

The study by \cite{vanDokkum2018_nat} was based on multi-slit spectroscopy of ten GCs, as no spectroscopic data were yet available for the stellar body of DF2. The stellar populations, GC kinematic association, and internal velocity dispersion of the galaxy were still unknown.

In this paper, we present the first spectroscopic observations of the stellar body of DF2 itself, obtained with the Multi-Unit Spectroscopic Explorer (MUSE) at the ESO Very Large Telescope (VLT). This integral field spectrograph allowed us to simultaneously obtain the spectra of seven associated GCs, two of which did not yet have a measured systemic velocity, as well as three newly discovered planetary nebulae (PNe). We used this dataset to consistently compare the receding velocities and stellar populations of the galaxy and its GC and PNe systems. We briefly present our data and analysis in Sect.~\ref{sec:data}, also describing our approach to extract the kinematics of the galaxy, GCs, and PNe. Results on the
kinematics are presented in Sect.~\ref{sec:results}, and then discussed in Sect~\ref{sec:discussion}. We briefly summarise and conclude in Sect~\ref{sec:summary}.
A companion paper focuses on constraining the detailed properties of the associated stellar populations \citep[Paper II,][]{Fensch2019}. 

Throughout this paper, magnitudes are given in the AB system. Unless stated otherwise, error bars denote 1$\sigma$ uncertainties, or 68\% confidence intervals.
%-------------------------------------------------------------------

\section{Data and analysis}
\label{sec:data}

\begin{figure*}
   \centering
   \includegraphics[angle=0,width=6cm]{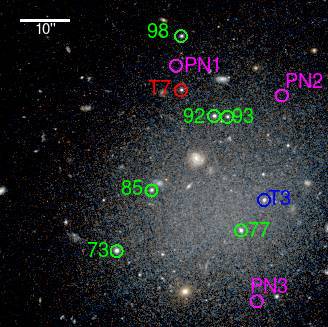}
   \includegraphics[angle=0,width=6cm]{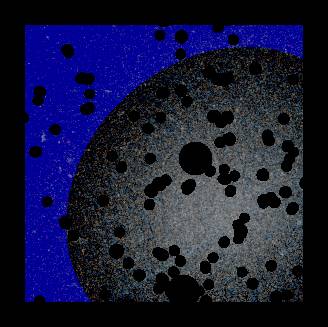}
   \includegraphics[angle=0,width=6cm]{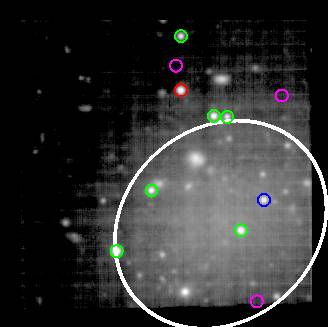}
      \caption{Left: Colour image of NGC1052-DF2 using F606W and F814 HST imaging over the MUSE field of view. Globular clusters and planetary nebulae are indicated. Green circles mark globular clusters identified in \citet{vanDokkum2018_nat}. The blue circle marks a GC candidate from \citet{Trujillo2018}, which we confirm to be a member of the galaxy based on its velocity, while the red circle marks a GC candidate which, due to its very low systemic velocity, likely is a Galactic foreground star. Purple circles mark PNe with systemic velocities consistent with the galaxy. Middle: Same image but with the mask superimposed (in black). Right: MUSE cube stacked using all channels up to 8950 $\AA$ that are not masked in our analysis (due to residuals of bright sky lines). The GCs, PNe, and galactic  region are indicated as before. The white ellipse marks the region within 1~\re, used in this work to extract the main galaxy spectrum. The blue region highlights the region used by \texttt{ZAP} to construct the set of sky eigen-spectra.}
         \label{fig:UDG}
   \end{figure*}

VLT/MUSE observations of NGC1052-DF2 were conducted via ESO-Director Discretionary Time (DDT) programmes (2101.B-5008(A) and 2101.B-5053(A), PI: Emsellem) in service mode during dark-time periods from July to November 2018. Seven observing blocks (OBs) were executed, amounting to a total of $\sim 5.1$h on-target integration time. Five OBs were observed with seeing between $\sim 0\farcs4 - 0\farcs6$ while the last two had seeing between $\sim 0\farcs7$ and $1\farcs0$. This allowed us to study the galaxy out to 1.5 effective radius and down to a surface brightness of nearly 26~mag~arcsec$^{-2}$ (I$_{814}$ band), as well as to extract properties of faint spatially unresolved GCs and PNe of DF2. Each OB was split into four individual on-target exposures with slight dithers and rotations to average out cosmic rays and patterns associated with the slicers. We deliberately offset the MUSE field with respect to the centre of the galaxy by $\sim 8\arcsec$ to probe a region where the surface brightness of the galaxy is several magnitudes fainter than at the centre (Fig.~\ref{fig:UDG}), and extract relevant spectral information for the sky background, as explained below. The first three OBs (July) had an "S O O S O O S" sequence (with S being offset skies, 180s each, and O being the object or galaxy exposed for 637s each) while the last four OBs (November) had an "S O O O O S" sequence, with the same 180s on-sky exposures and slightly longer 672s exposures on target. Offset skies were meant to serve as a rescue plan for the sky subtraction, but were ultimately used for consistency checks only.

\subsection{Reduction steps}
\label{sec:reduction}

All MUSE OBs were reduced using the latest MUSE esorex pipeline recipes (2.4.2), all wrapped within a dedicated python package (pymusepipe, available via {\tt github}). The reduction follows the standard steps (bias, flat, wavelength calibration, line spread function, illumination correction), and produces the so-called pixel tables which are corrected for barycentric motion, combined and resampled after alignment using the $\tt{exp\_combine}$ MUSE pipeline recipe.  

The MUSE pipeline includes a wavelength calibration based on arc lamp exposures that were taken during the day, after correcting for the difference in temperature. The reported accuracy is around 0.03~\AA\ (rms). The pipeline also corrects each exposure for any residual global shifts using the detected sky lines. We performed a consistency check using the positions of the sky lines in the object frames themselves, confirming that the resulting wavelength calibration is accurate within 2~\kms, and is uniform over the field of view of MUSE.

We identified and subtracted the sky background using the Zurich Atmosphere Purge (\texttt{ZAP}) software package \citep{Soto2016}, a principal component analysis-based sky subtraction tool that is optimized for Integral-Field Units (IFUs). The principal components (eigen-spectra) are derived from the outermost regions of the MUSE object cube, where the sky is dominant (see the blue region in the middle panel of Fig.~\ref{fig:UDG}), after masking foreground and background sources. We also experimented by letting \texttt{ZAP} set up the eigen-spectra on the external sky exposures (and then fitting them to the object cube), but found an inferior performance compared to exclusively using the sky-dominated region in the object stack due to time variations in the sky background and the significantly shorter exposure time. Even though this is an area to investigate further, we speculate that the flat-fielding of MUSE is sufficiently good that sky regions that are observed at the exact same time, yet falling on a different place on the detector, are preferred over the offset exposures. In Appendix \ref{app:skysubtraction} we describe the tests we performed to ensure that the source spectra are not affected by \texttt{ZAP}. 

\subsection{Detection and spectral extraction}

We extracted spectra for the globular clusters studied in \citet{vanDokkum2018_nat} (green circles in Fig.~\ref{fig:UDG}), as well as the globular cluster candidates reported in \citet{Trujillo2018} (blue and red circles in Fig.~\ref{fig:UDG}), which fall within the MUSE field of view.
Furthermore, in order to detect possible planetary nebulae around DF2 we cut the cube into three slabs of $8~\AA$ centred on the redshifted wavelength of the [O~\textsc{iii}] doublet and H$\alpha$ emission lines, from which we subtracted three slabs of the same width in spectrally close featureless regions. Three sources with unambiguous detection of the [O~\textsc{iii}] doublet and H$\alpha$ line were found. All GCs and PNe positions are shown in Fig.~\ref{fig:UDG}. 

The spectra were extracted with a Gaussian weight function, whose full width at half maximum (FWHM,
thereafter) is chosen to be $\sim0\farcs8$ to approximately match the point spread function (PSF), providing a signal to noise (S/N)-optimised extraction. The apertures were truncated at $2\farcs4$ to avoid contamination from neighbouring sources. The background was probed locally with identical apertures in eight nearby locations that do not overlap with identified sources. In each channel, we then subtracted the median value of the eight background regions from the source flux (which includes the background). The dominant source of uncertainty thus originates from the scatter in the background spectrum values.

After the sky subtraction performed using \texttt{ZAP}, we extracted the signal from DF2 in two different ways. The first extraction method, providing a global spectrum, uses the whole MUSE field, and adopts the F814W Hubble Space Telescope (HST) image\footnote{The HST data are available in the Mikulski Archive for Space Telescopes (MAST; http://archive.stci.edu), under program ID 14644.} as optimal weight function. For this we registered the F814W image to the same 0.2$\arcsec$/pix grid as the MUSE cube, convolved the PSF to match the PSF of the MUSE cube, and identified and masked all compact sources that are detected on this deep broadband image. The mask is shown in Fig.~\ref{fig:UDG}. By performing the extraction in each 1.25$\AA$ channel separately, we obtained a global DF2 spectrum with a S/N peaking at $\sim$75~pix$^{-1}$ at 7000$\AA$, and dropping to $\sim$40~pix$^{-1}$ towards the blue and red end of the spectral range.

   \begin{figure*}
   \centering
   \includegraphics[angle=0,width=18cm]{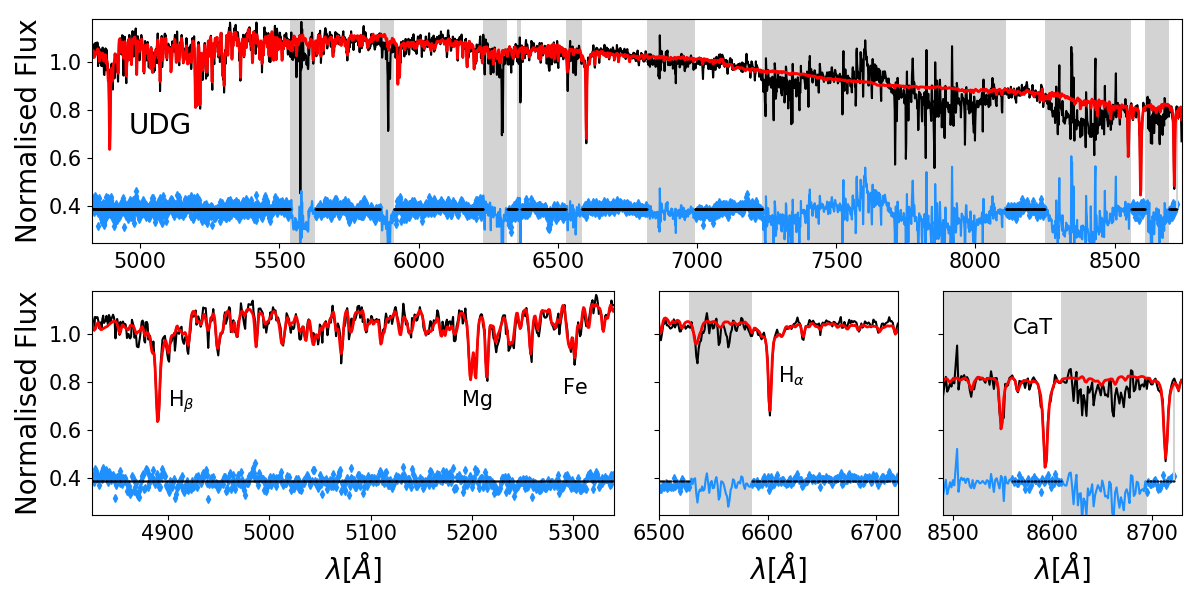}
      \caption{Spectrum of DF2 integrated over an aperture of 1~$R_{\mathrm{e}}$ (black). The pPXF best fit
      is shown as a red line, while the residuals are shown as blue dots. 
       The top panel illustrates the full MUSE spectral coverage, while the bottom three panels present
      zoomed versions around the main stellar absorption lines. Grey regions indicate wavelength ranges that have been excluded from pPXF fitting because of large sky line residuals.}
         \label{fig:UDGspectrum}
   \end{figure*}

We also performed extractions within several elliptical apertures, corresponding to $0.5$, $1.0,$ and $1.5$ times the effective radius \re\ as measured along the major axis. We used morphology parameters as measured in \citet{vanDokkum2018_nat}: the galaxy centre is taken at RA=40.44542$^{\circ}$, Dec=$-$8.40333$^{\circ}$, with axis ratio $b/a=0.85$, and a position angle of $-$48\degree (measured from the north towards the east) from the HST images. The spectrum measured within 1~\re\ is presented in the top panel of Fig.~\ref{fig:UDGspectrum}. 

We finally built an adaptive spatially binned version of the MUSE datacube, following a Voronoi binning scheme \citep{CappellariCopin} with a minimum target S/N of 15. That binning scheme should reveal any velocity gradient over the MUSE field of view, as well as potential trends along the photometric minor or major axes.

\subsection{Stellar kinematics}
\label{sec:measureveldisp}

To measure kinematics of the globular clusters and the galaxy stellar body, we first construct a wavelength-dependent mask to discard regions where groups of bright sky lines significantly affect the stellar kinematics. The mask is shown by the greyed-out region in Fig.~\ref{fig:UDGspectrum}; we use the same mask for the galaxy as for the globular clusters. We also produce wavelength-dependent resolution profiles as in \cite{guerou17} for each individual extracted aperture (Voronoi bins and elliptical apertures) by measuring the width of the sky emission lines directly on the original (deep) merged datacube, and fitting the measurements via second-degree polynomials: these are comparable to the "UDF10" profile used in \cite{guerou17} (see Appendix~\ref{app:resprofiles}).

We extract kinematics mainly with the penalised pixel fitting routine, pPXF, whose principle is described in \cite{Cappellari2004} \citep[see also][]{Cappellari2017} and is available as a set of python routines.\footnote{See \texttt{http://www-astro.physics.ox.ac.uk/\textasciitilde{}mxc/software/.}} The routine pPXF provides the best-fit spectrum (minimising the Chi-square of the residuals) assuming a linear combination of a given set of spectral templates. We use two empirical stellar libraries as spectral templates, namely eMiles \citep{Vazdekis2016, Roeck2016} and Pegase-HR \citep{LeBorgne2004}. The eMiles library has the advantage that it covers the full wavelength range of MUSE spectra, a relatively large set of single stellar populations (SSPs) templates built on empirical stellar spectra, with a medium spectral resolution \citep[2.54$\,\AA$~FWHM, see][]{beifiori11}. We use a set of 250 eMiles SSP spectra with a Kroupa initial mass function (IMF), metallicities [Fe/H] ranging from solar down to -2.32, and ages from 63~Myr to 17.78~Gyr. Pegase-HR has a significantly higher spectral resolution, while only covering wavelengths up to a bit beyond the H$_{\alpha}$ line. We include 408 Pegase-HR spectra with Kroupa IMF, metallicities from solar to -2.0, and ages from 1~Myr to 16~Gyr. We present results from both libraries, also testing for the robustness of our results by restricting the fitted spectral range.

\subsubsection{Velocities}
\label{sec:vel}

We derive stellar kinematics with pixel fitting via pPXF and a set of stellar templates. All template spectra are convolved to the MUSE resolution using the polynomial approximation associated with each aperture or bin (see Appendix~\ref{app:resprofiles}). All templates (and MUSE) spectra were then rebinned in logarithm of the wavelength and fed into pPXF. The first step is to derive an optimal (best-fit) stellar template using a selected set of eMiles (or Pegase-HR) spectra, constraining the metallicity to be lower than solar. One optimal template is derived using the integrated (1~\re) DF2 spectrum, and another template is computed for the clusters via a stacked GC spectrum. For this first step, we allow for high order multiplicative and additive Legendre polynomials of the 14th and 12th degree, respectively, to correct for differences in the flux calibration and to go beyond limitations of the stellar library.

We then run pPXF to extract stellar velocities, using the respective optimal templates as a single stellar template for the fitting process, still allowing for multiplicative (resp. additive) polynomials. The uncertainties are estimated by performing 1000 realisations of the full spectrum using a wild bootstrap methodology \citep{Wu1986} as in \citet{Kacharov2018}.

We performed several checks to ensure that these velocity measurements are robust. The results are independent of the way the background was subtracted from the MUSE cube (even when ZAP is not used, cf.~Appendix~\ref{app:skysubtraction}). Also, the degrees of the Legendre polynomials do not significantly affect the results; we checked values between 5 and 14 in steps of 3. The stellar velocities are not significantly affected by the choice of the optimal template (e.g. when we interchange the globular cluster and the galaxy template), or even the assumed wavelength-dependent spectral resolution (see Sect.~\ref{sec:disp}), with differences of the order of 0.5~\kms. We also split the spectra into a blue and red part (centred on 7000$\AA$): all recovered receding velocities are consistent within uncertainties.

Furthermore, we performed a consistency check of the measured systemic velocities that is independent of the fitting method used; we processed all MUSE spectra via a cross-correlation (CC) method, which is widely used to provide receding velocity measurements of faint galaxies \citep{tonrydavis79}. The CC velocities are also reported with respect to the eMiles spectral library, and for the PNe with respect to a template spectrum generated using optical emission line data provided with the automated line fitting algorithm (ALFA) code \citep{wesson16}. All velocity measurements obtained with pPXF and cross-correlations are consistent within less than 1-$\sigma$ (cf.~Table~\ref{table:sysvels}). In the following analysis we adopt the values obtained with pPXF, unless explicitly stated otherwise.

The systemic velocity of the galaxy is measured with the best-fit DF2 template. We used the same reference template when measuring the systemic velocities in each of the Voronoi bins. We measured the statistical uncertainties in the same way as for the globular clusters, and performed the same robustness tests as before. In all cases, the systematic uncertainties due to different assumptions in the analysis are smaller than the quoted statistical uncertainties. As a note, the use of the Pegase-HR library induces slightly lower velocities, with a systematic difference of about 2.0~\kms, than with eMiles, using the same procedure. This is a systematic shift in both the galaxy velocity field and the velocities measured for the GCs, and would only affect the relative velocities with respect to the PNe: such an offset does not affect any of the conclusions drawn in this work.

\subsubsection{Velocity dispersions}
\label{sec:disp}

We expect stellar velocity dispersions for the galaxy and clusters typically between 5 and 35~\kms, which thus represent a substantial fraction of the spectral resolution of the MUSE datacube ($\sigma_{inst}$ going from $\sim 35$ to 80~\kms). To estimate the velocity dispersion of DF2 and the clusters, which are derived relative to convolved templates, we therefore need precise determinations of the spectral resolution, either parameterised via a wavelength dependent line spread function (LSF), or further approximated with a Gaussian function (via its FWHM), for both the MUSE datacube and for the reference stellar templates.

As mentioned above, we use wavelength-dependent resolution profiles given by a polynomial approximation FWHM$(\lambda)$, using the sky lines in each aperture or bin spectrum (before sky subtraction and without barycentric correction). The lower the assumed spectral resolution, the higher the resulting estimate for the stellar velocity dispersion. We tested the dependence of our results introducing a perturbation of the resolution profiles consistent with the measured scatter in the FWHM of the sky lines: all dispersion values are consistent, within the one-sigma uncertainties. It is interesting to note that using a fixed single resolution profile for all Voronoi bins leads to a local increase of the dispersion values in the south-west corner of the field of view, corresponding to the region where the sky lines have slightly narrower FWHMs (see Appendix~\ref{app:resprofiles}). When using our profiles, this local increase disappears and all values are consistent with a flat dispersion field (within one-sigma uncertainties). This suggests that the small measured variation of the spectral resolution over the field of view is relevant. We have also tested that using a Gaussian approximation for the LSF or performing a direct and full LSF convolution provides consistent results. In the following, we will thus quote results from our measured individual resolution profiles.

%-------------------------------------------------------------------
\section{Results}
\label{sec:results}

\subsection{Velocities of globular clusters and planetary nebulae}

We measured systemic velocities for six of the clusters that overlap with the sample presented in \citet{vanDokkum2018_gc} (see Table~\ref{table:sysvels}). Globular cluster candidates T3 and T7 are reported in \citet{Trujillo2018} (and called GCNEW3 and GCNEW7 there) on the basis of their small apparent size and colours. One of those candidates, T3, has a systemic velocity consistent with that of the other globular clusters. The other candidate, T7, has a velocity consistent with a Galactic foreground star: inspecting its MUSE spectrum, we hypothesise that it is an early G-type star. 

We compared the measured velocities with those reported in \citet{vanDokkum2018_nat}, and found a systematic offset of $12.5_{-2.5}^{+2.6}$~km.s$^{-1}$ when using the eMiles library as a reference (or 14.5~km.s$^{-1}$ when using Pegase-HR templates). This includes a shift of 5.4~km.s$^{-1}$ due to the different applied redshift-velocity transformation ($\mathrm{c}\,z$ versus $\mathrm{c}\ln(1 + z)$ for the present paper\footnote{See \cite{Baldry2018}.}). Still, it is more than we expect given the estimated accuracy of our wavelength calibration (cf.~Sect.~\ref{sec:reduction}). The source of this systematic discrepancy is unclear. During the kinematic studies presented in this paper, we explicitly include this systematic offset as a prior in our likelihood when combining measurements from the two studies.

We also measured systemic velocities for the three PNe, facilitated by their [O~\textsc{iii}] doublet and H$\alpha$ emission lines, which are detected at high significance \citep[see][Paper II]{Fensch2019}. The measurements and associated errors are reported in Table~\ref{table:sysvels}.

\begin{table*}
\caption{Positions and systemic velocity measurements of the unresolved, point-like sources and DF2 itself (using the eMiles library). Both results from the fitting method (pPXF) and the cross-correlation (CC) method are given. For the galaxy and the GCs, the S/N is measured in the wavelength range [$6600\AA-6800\AA$] in pix$^{-1}$ (one pixel corresponds to a 1.25$\AA$ channel). For the PNe the S/N is given for the [O~\textsc{iii}] $\lambda5007$ line. First (second) errors on the measured velocities represent statistical (systematic) uncertainties. The latter are the most extreme values obtained given a range of tested templates, background subtraction methods, and polynomial degrees of the Legendre function that was fitted to the continuum. For completeness, we augment the table with five GCs that were not covered by our MUSE observations, but for which velocities were reported by vD18 and for which we take the coordinates in \citet{Trujillo2018}.}   
\label{table:sysvels}      
\centering          
\begin{tabular}{l l l l l l l l}     % 4 columns 
\hline\hline       
Source & RA & Dec & S/N & V (pPXF) & V (CC) & vD18 ($c \cdot z$)\\
& [deg] & [deg] & pix$^{-1}$ & [km/s] & [km/s] & [km/s] \\
\hline                    
DF2$(<{R_{\mathrm{e}}})$ &  40.44542 & -8.40333 & 75.1 & $1792.9_{-1.8-1.3}^{+1.4+0.2}$ & $1794.5_{-1.5-1.7}^{+1.7+0.4}$ & -\\
GC73 &  40.45093 & -8.40503 & 47.6 & $1803.8_{-3.0-0.7}^{+2.8+3.3}$ & $1803.9_{-2.2-0.8}^{+2.2+0.7}$ & $1814_{-3}^{+3}$\\
GC77 &  40.44395 & -8.40390 & 30.7 & $1792.6_{-4.7-11.6}^{+4.4+1.8}$ & $1787.9_{-2.9-1.4}^{+3.6+1.2}$ & $1804_{-6}^{+6}$\\
GC85 &  40.44896 & -8.40166 & 27.1 & $1786.3_{-4.8-6.1}^{+4.3+1.9}$ & $1786.1_{-2.2-0.8}^{+3.2+1.5}$ & $1801_{-6}^{+5}$\\
GC92 &  40.44544 & -8.39753 & 28.9 & $1775.5_{-4.2-4.7}^{+6.2+2.8}$ & $1772.1_{-3.6-4.1}^{+2.2+2.2}$ & $1789_{-7}^{+6}$\\
GC93$^{\mathrm{a}}$ &  40.44469 & -8.39759 & 16.0 & $1819.1_{-6.8-10.5}^{+7.7+21.1}$ & $1818.6_{-6.9-4.4}^{+5.5+1.1}$ & -\\
GC98$^{\mathrm{b}}$ &  40.44728 & -8.39310 & 18.0 & $1786.3_{-11.6-5.7}^{+7.4+22.9}$ & $1782.7_{-8.3-4.1}^{+9.4+3.7}$ & $1784_{-10}^{+10}$\\
T3$^{\mathrm{c}}$ &  40.44265 & -8.40221 & 21.6 & $1788.7_{-10.9-47.7}^{+16.9+16.3}$ & $1794.1_{-6.4-10.1}^{+5.9+5.1}$ & -\\
T7$^{\mathrm{c}}$ &  40.44729 & -8.39610 & 30.0 & $22_{-16}^{+19}$ & $52_{-44-15}^{+15+15}$ & -\\
PN1 &  40.44758 & -8.39473 & 20.3 & $1790.8_{-1.8}^{+2.0}$ & $1790.5_{-1.9-1.1}^{+2.8+1.4}$ & -\\
PN2 &  40.44164 & -8.39639 & 6.6 & $1786.4_{-5.9}^{+7.2}$ & $1782.3_{-7.5-0.6}^{+9.9+2.6}$ & -\\
PN3 &  40.44305 & -8.40783 & 5.8 & $1766.4_{-7.1}^{+10.2}$ & $1763.3_{-5.0-3.5}^{+6.9+1.5}$ & -\\
GC39 &  40.43779 & -8.42358 & - & - & - & $1818_{-7}^{+7}$\\
GC59 &  40.45034 & -8.41596 & - & - & - & $1799_{-15}^{+16}$\\
GC71 &  40.43807 & -8.40638 & - & - & - & $1805_{-8}^{+6}$\\
GC91 &  40.42571 & -8.39832 & - & - & - & $1802_{-10}^{+10}$\\
GC101 &  40.43837 & -8.39120 & - & - & - & $1800_{-14}^{+13}$\\
\hline                  
\end{tabular}
\begin{list}{}{}
\item[$^{\mathrm{a}}$] Observed in \citet{vanDokkum2018_nat} but no measured velocity due to too low S/N in their spectrum.
\item[$^{\mathrm{b}}$] Re-measured value reported
in \citet{vanDokkum2018_vel}.
\item[$^{\mathrm{c}}$] Globular cluster candidate reported in \citet{Trujillo2018}.
\end{list}
\end{table*}

\subsection{Kinematics and dynamics of DF2 stellar body}
\label{sec:udgkinematics}

Following the methodology as outlined in Sect.~\ref{sec:vel}, we measured a systemic velocity of $1792.9_{-1.8}^{+1.4}$ for the MUSE spectrum extracted within the effective radius. We measured identical values on the spectrum extracted from the full frame (weighted by the broadband image), like the one extracted within \re. This value is fully consistent with the mean velocity of the globular clusters (cf.~Table~\ref{table:sysvels}). The measured velocity provides the rest frame of the whole system \citep[which previous studies had to solve for in their likelihood, e.g.][]{Martin2018,Laporte2018}. We use this systemic velocity when re-measuring the velocity dispersion of the GC system in Sect.~\ref{sec:discussion}.
We also find that spectra measured within 0.5 and 1.5~\re\ apertures both provide consistent velocity values, within $1\,\sigma$ uncertainties (of the order of 1.5~\kms).

\begin{figure}
   \centering
   \includegraphics[angle=0,width=9cm]{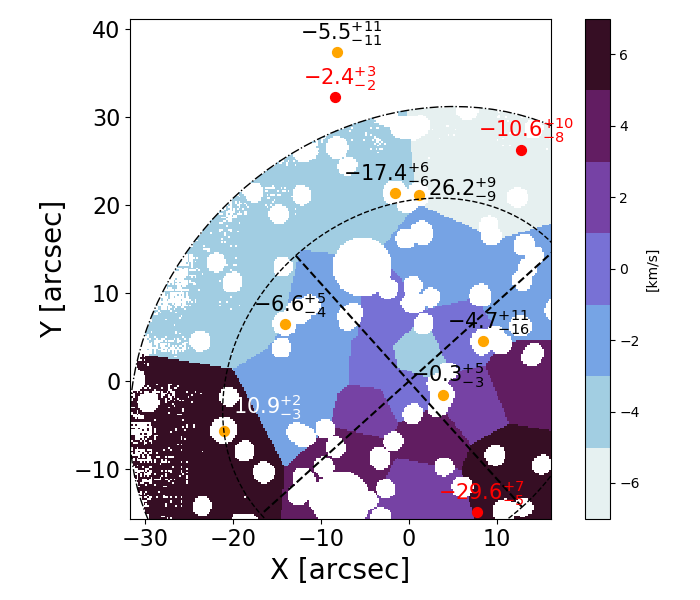}
      \caption{Stellar velocity field in Voronoi bins of DF2. Orange dots and associated measurements provide the positions and systemic velocities of the seven globular clusters that are members of the galaxy, while the red dots denote the positions and velocities of the three discovered PNe. The velocities are indicated with respect to the systemic velocity of 1792.9~\kms. The two ellipses correspond to 1.0~$R_{\mathrm{e}}$ and 1.5~$R_{\mathrm{e}}$, respectively. The X- and Y axes denote the angular difference with respect to the centre, RA=40.44542$^{\circ}$, Dec=$-$8.40333$^{\circ}$, in the western and northern direction, respectively.}
         \label{fig:stellarVmap}
\end{figure}

\begin{figure}
   \centering
   \includegraphics[angle=0,width=9cm]{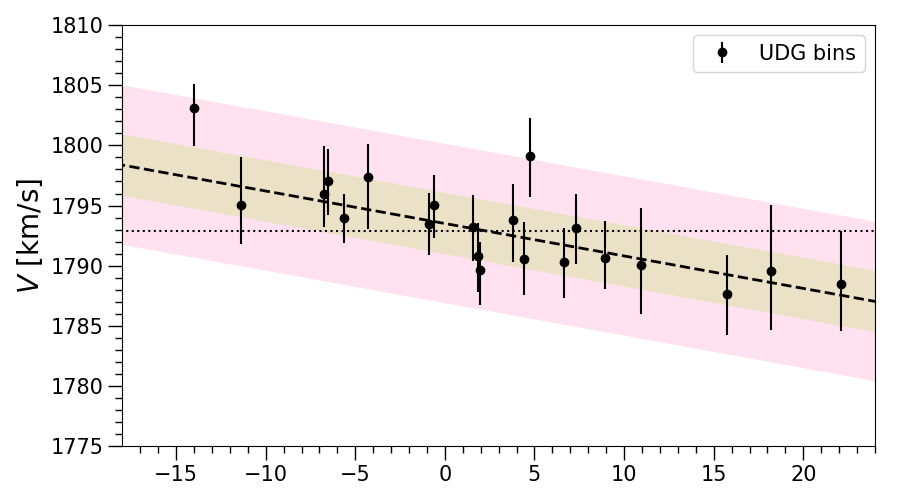}
   \includegraphics[angle=0,width=9cm]{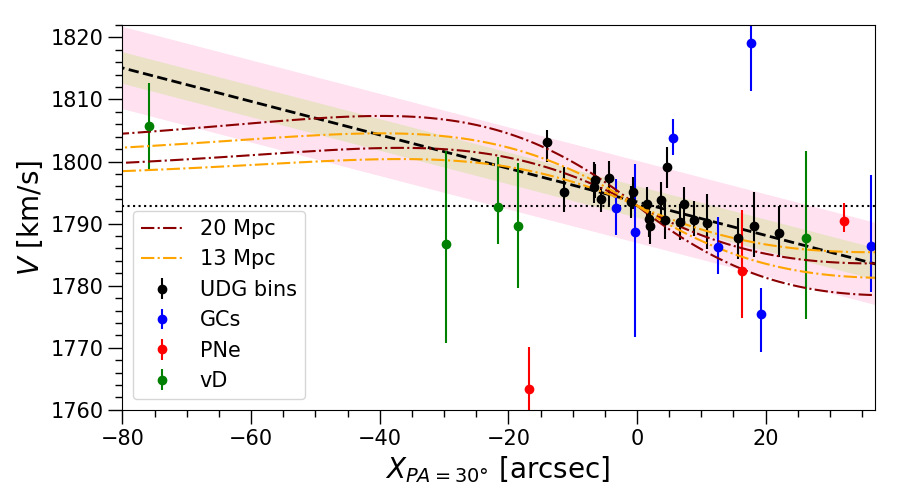}
      \caption{Stellar velocity field trend shown along a position angle of $+30\degree$ for the DF2 Voronoi bins (in black). The dotted horizontal line shows the systemic velocity derived from the 1~\re\ elliptical aperture. The dashed line shows the best linear fit on the DF2 bins, with the residual scatters (1 and 2.6~$\sigma$) scatter from the DF2 bins alone in shaded green and red. Measured clusters and PNe are respectively shown as blue and red points. The top panel shows a zoom within the MUSE field of view. The bottom panel includes additional clusters measured by \citet{vanDokkum2018_nat}, as well as reference circular velocity profiles from a S\'ersic model at a distance of 20 or 13~Mpc with M/L$_V = 2$ , and for two different inclination angles (35 and 90$^\circ$, see text).}
         \label{fig:gradV}
\end{figure}

To be able to interpret the measured velocity dispersion of the clusters as a proxy for the dynamical mass of the system, it is important to be aware of any rotational component to the velocity field of the galaxy. \citet{vanDokkum2018_nat} did not find any evidence for rotation (their Extended Data Fig. 5), when they inspected the globular cluster velocities. 
Figure~\ref{fig:stellarVmap} presents the MUSE stellar velocity field as derived in the adaptive Voronoi bins of DF2. All obtained bin velocities are within $\pm 6$~\kms of the mean systemic velocity, except for one outer bin at $+10$~\kms, while uncertainties are roughly $\pm 3$~\kms (1~$\sigma$). 
The MUSE stellar velocity map exhibits a clear trend with positive velocities roughly towards the south and negative velocities mostly towards the north of the centre. The split between (relative) negative and positive velocities seems to be on either side of the photometric major axis of the galaxy. The strongest detected gradient, evaluated by performing linear fits of the measured velocity with respect to a rotated axis, is along a position angle PA$=30\degree$ (anti-clockwise from north-up axis in Fig.~\ref{fig:stellarVmap}), with similar slopes in the PA range [$10\degree$,$47\degree$], which includes the photometric minor axis at PA$=42\degree$: the linear slope is $2.8 \pm 0.9$~\kms per 10\arcsec, as illustrated in Fig.~\ref{fig:gradV}, with potential large systematic uncertainties due to the shape of the bins and the limited field of view. This measurement is inconsistent with an oblate rotator, while mimicking a system with prolate-like kinematics.

In Fig.~\ref{fig:gradV}, we added the above-mentioned linear relation and the residual stellar velocity scatter (1 and 2~$\sigma$) within the Voronoi bin set. We also derived the circular velocity profiles expected from a simple S\'ersic stellar distribution. These are derived from building full three-dimensional models for NGC1052-DF2 via the multi-Gaussian expansion formalism \cite[ MGE][]{Emsellem1994}. A projected S\'ersic $n=0.6$ system following morphological parameters (e.g. axis ratio $q=0.85$) as in \cite{vanDokkum2018_nat} was fitted with an MGE model and deprojected using different inclination angles $\theta$ from edge-on ($90\degree$) to close to face-on ($35\degree$). We assumed two generic distances as discussed in the literature, namely 20 and 13~Mpc \citep{vanDokkum2018_nat, Trujillo2018}, and used a reference stellar mass-to-light ratio of $M_{\rm star}/L_V =2$ \citep{vanDokkum2018_nat}. These profiles are meant as simple guidelines for the measured stellar velocity gradient.

Most GCs and PNe (respectively represented as blue and red symbols in Fig.~\ref{fig:gradV}) are consistent with the general uncovered velocity gradient: only one cluster seems to stand out, namely GC93 with its measured relative radial velocity of $\sim +26$~\kms, very significantly above the main relation, and one out of three PNe (PN3) appears to be an outlier with $\sim -26$~\kms with respect to systemic velocity. This will be briefly discussed further in Sect.~\ref{sec:revdisp}.


\subsection{Stellar velocity dispersion of DF2}
\label{sec:dispUDG}

We derived a nominal stellar velocity dispersion value of $\sigma_{\rm{DF2}\star}$(\re)$ = 10.8_{-4.0}^{+3.2}$~\kms\ (see Sect.~\ref{sec:disp}) using the eMiles stellar library, with the 95\% confidence interval providing an upper limit of 17.4\kms. Similar values are found for 0.5 and 1.5~\re\ with $9.4_{-4.7}^{+3.6}$ and $10.3_{-5.9}^{+4.3}$~\kms, respectively. Such a dispersion measurement relies on a few narrow absorption lines, including mostly \Ha, \Hb, Mg, and Fe
lines between 5000 and 5500~\AA, the CaT triplet around 8500~\AA\ (see Fig.~\ref{fig:UDGspectrum}). 

Extensive tests meant to pin down the effects of various assumptions (spectral domain, resolution profile, degree of additive polynomials) show that it is sensitive to associated changes. Such instabilities are mostly connected with the difficulty to extract small dispersion values while systematics from the assumed LSF and the intrinsic broadening of the relevant absorption lines in different templates (which is linked to the evolutionary stage of the stellar population) are present. We therefore tested the different template libraries (Sect.~\ref{sec:measureveldisp}) and different (narrower) spectral regions, focusing on for example the blue part of the spectra (blueward of 7000~\AA), or solely around the CaT region where the MUSE spectral resolution is highest (but very close -- within 0.1~\AA\ -- to the eMiles library resolution).

When using the Pegase-HR library, which only covers the region bluewards of 7000~\AA, we get a distribution of values with a peak at the lowest boundary (set to 1~\kms, and not 0, for numerical reasons), still with more than half of the values above 3~\kms. The one-sigma and two-sigma upper boundaries are 12.7 and 17.9~\kms respectively, encompassing the nominal value. Within the same (bluer) region, eMiles provides a value of $\sigma_{\rm{DF2}\star}$(\re)$ = 12.1_{-3.7}^{+3.4}$~\kms. When using eMiles, this time restricting the fit to the CaT region, we find $\sigma_{\rm{DF2}\star}$(\re)$ = 5.4_{-4.4}^{+7.5}$~\kms, hence lower but still consistent with the above-mentioned values.

We also derived individual values of the stellar velocity dispersion for each Voronoi bin: the dispersion map is consistent with being flat (within the one-sigma uncertainties), with values going from 5.5 to 12.6~\kms\ and a one-sigma upper limit of 23~\kms: this is a significantly broader range accounting for the lower S/N ratio of individual bins.

\begin{figure}
   \centering
   \includegraphics[angle=0,width=9cm]{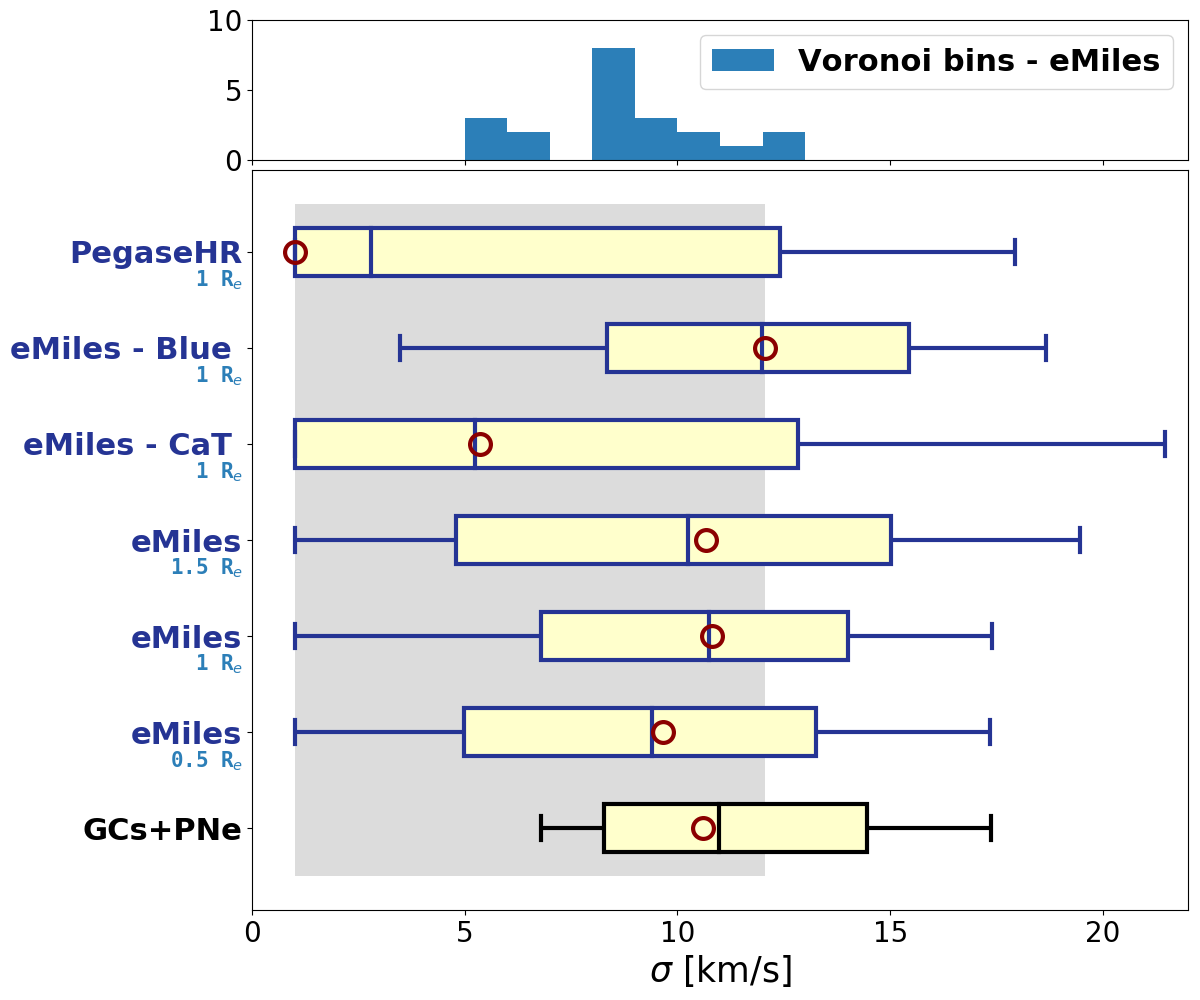}
      \caption{Stellar velocity dispersion values as measured from the MUSE spectra. The bottom panel shows
      individual measurements for the three extracted apertures (0.5, 1.0, and 1.5~Re) mentioned in the text, when using
      different template libraries (eMiles or Pegase-HR) and different spectral coverage (around the Calcium triplet, or
      a "Blue" region - below 7000~\AA). The red circles are the actual measurements, while the vertical bar
      shows the medians of the noise realisations. The extent of the boxes and whiskers correspond to the [16,84] and [2.5,97.5]
      percentiles, respectively. The line at the bottom labelled "GCs + PNe" indicates the values obtained via discrete
      tracers (see Sect.~\ref{sec:revdisp}). The grey area illustrates the systematics by encompassing the
      various measurements. The top panel presents an histogram of the individual measurements for the 21 Voronoi
      bins, using the eMiles stellar templates. 
              }
         \label{fig:dispersions}
\end{figure}

In Fig.~\ref{fig:dispersions} we illustrate these measurements, also providing values for the medians, the 68 and 95\% confidence intervals of the realised distributions. To sum up these various tests, we find a nominal value of $\sigma_{\rm{DF2}\star}$(\re)$ = 10.8_{-4.0}^{+3.2}$~\kms, with values ranging from 5 up to 12~\kms\ when solely using the eMiles library, and a two-sigma upper limit of $\sim 21$~\kms\ when including all systematics. We note that the measured dispersion includes both ordered and random motions, hence also the rotational component within the considered aperture. We finally note that the dispersion value within a smaller field of 0.5~\re\ quoted by \cite{Danieli2019} is consistent with our above-mentioned value. 

%-------------------------------------------------------------------
\section{Discussion}
\label{sec:discussion}

\subsection{Revised dispersion measurement}
\label{sec:revdisp}

We used the set of globular clusters for which we obtained reliable systemic velocities (including two newly measured clusters), the three PNe, and our established central velocity for the entire system to revisit the velocity dispersion measurement of this system. We followed a similar approach to \citet{Martin2018} and \citet{Laporte2018} to derive the dispersion from a likelihood fit to the kinematic data.

Given that their receding velocities are close to that of the main DF2 galaxy body, we do assume that all GCs and PNe are part of the galaxy, and we write the likelihood function as 
\begin{align}
   \mathcal{L}= \prod_{i=1}^{N_{GC,PN}=15} \frac{1}{\sqrt{2\pi}\sigma_{\mathrm{obs}} } \exp\Bigg(-0.5\bigg(\frac{v_{i}-C_{i}-v_{\mathrm{sys}}}{\sigma_{\mathrm{obs}}}\bigg)^2\Bigg) \ ,\\
   \mathrm{with}\,\,\, \sigma_{\mathrm{obs}}^2=\sigma_{\mathrm{int}}^2+\sigma_{\mathrm{err}}^2.
\end{align}
Here, $\sigma_{\mathrm{err}}$ is the measurement uncertainty, $\sigma_{\mathrm{int}}$ is the intrinsic dispersion we are seeking, and $C_{i}$ accounts for the velocity calibration offset between our study and that of \citet{vanDokkum2018_nat}. Its value is therefore either 0 (when using tracers measured with MUSE), or an offset value, $\delta$, when we use tracers exclusively reported by \citet{vanDokkum2018_nat}. We measure $\delta$ as the offset that minimises the least squares statistic $\sum_i^N (V_{i,\mathrm{vD18}}-\delta-V_{i,\mathrm{MUSE}})^2/(eV_{i,\mathrm{vD18}}^2+eV_{i,\mathrm{MUSE}}^2)$, where $V$ and $eV$ are the velocities and velocity uncertainties from both studies. We sum over the five GCs that are observed in common. For the measurements performed with pPXF, we find $\delta=11.1\pm3.0$ ($\delta=12.3\pm2.7$ km/s for the cross-correlation velocities).
Since our measurements are more precise than those reported in \citet{vanDokkum2018_nat}, we exclusively use our measurements when available (i.e.~when covered by the MUSE field of view). The velocity of the system as a whole is $v_{\mathrm{sys}}$. We adopt a uniform prior of $0 < \sigma_{\mathrm{int}} < \, 30$~\kms, and a Gaussian prior for $v_{\rm sys}$ based on our MUSE kinematic measurements of the galaxy stellar body (cf.~Table~\ref{table:sysvels}).

\begin{figure}
   \centering
   \includegraphics[angle=0,width=9cm]{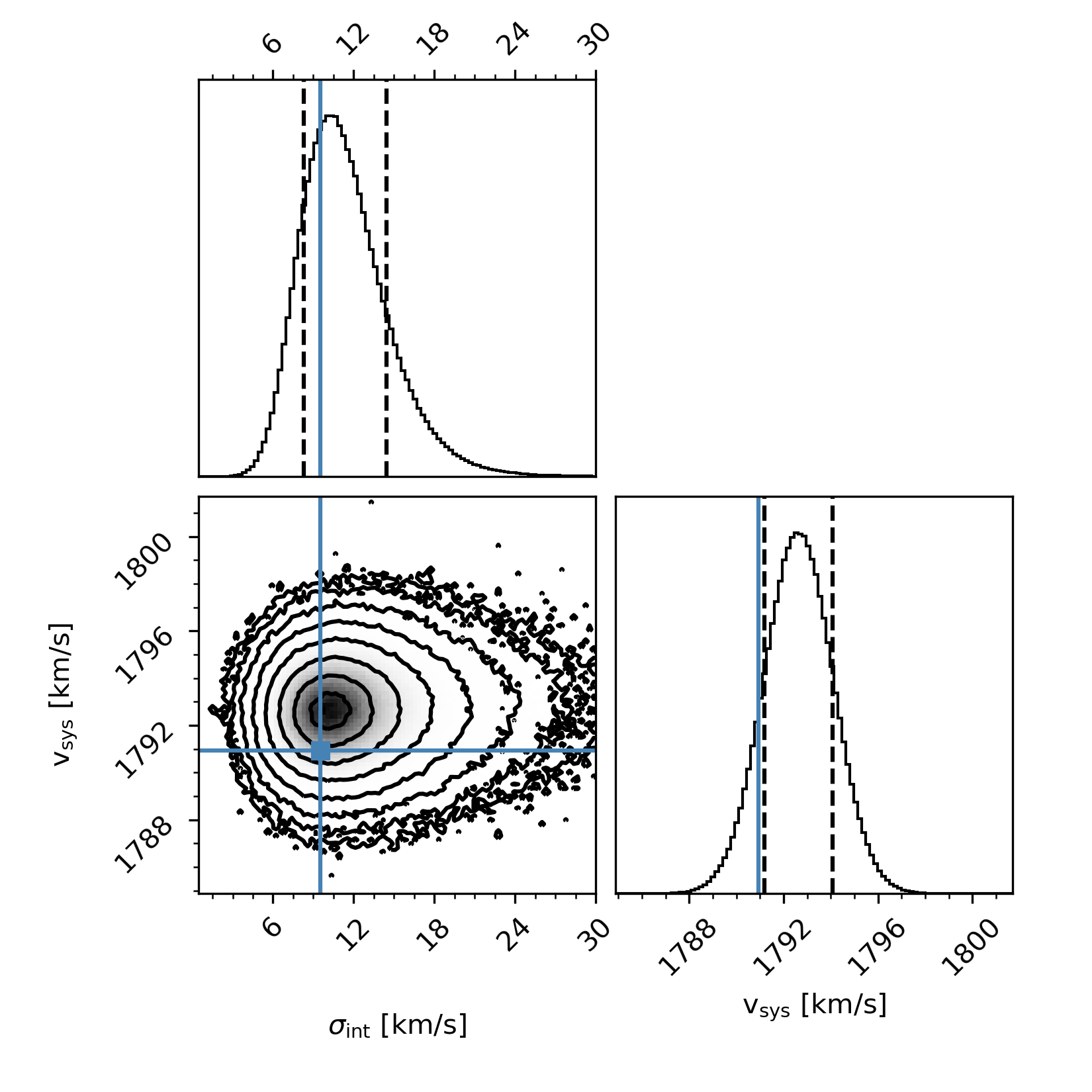}
      \caption{Inference on systemic velocity and internal velocity dispersion using discrete kinematical tracers. The blue point is the previous measurement \citep[][shifted by the velocity calibration bias]{Martin2018} before $v_{sys},$ the velocities of two additional GCs and three PNe, were measured.
              }
         \label{fig:mcmcresult}
\end{figure}

We marginalize over $\delta$ by drawing from a Gaussian distribution with $\mu$ and $\sigma$ according to the velocity calibration offset reported above. We draw 150 times from this distribution and for each draw explore the posterior with ten walkers using \texttt{emcee} \citep{emcee}. The combined posterior is shown in Fig.~\ref{fig:mcmcresult}. As a consistency check we run the Markov Chain Monte Carlo (MCMC) over the old dataset (with uniform prior on $1750\,\mathrm{km/s}<v_{\mathrm{system}}<1850\,\mathrm{km/s}$, and with the old velocity measurement of GC98) and find a best-fit velocity dispersion of $\sigma_{\mathrm{int}}=10.2^{+7.0}_{-2.5}$ km/s, which is consistent with \citet{Martin2018} and \citet{Laporte2018}. Folding in the additional two globular clusters, three PNe, and constraining the velocity of the system to the galaxy mean velocity gives a similar velocity dispersion of $\sigma_{\mathrm{int}}=10.6^{+3.9}_{-2.3}$ km/s. This assumes that PNe and GCs are equally good tracers of the potential and are all drawn from distribution functions with comparable line-of-sight dispersions. Velocities from the cross-correlation method yield $\sigma_{\mathrm{int}}=13.0^{+4.3}_{-2.1}$ km/s. We note that the results above are not corrected for small-sample bias as per \citet{Laporte2018}. With the current uncertainties, this would increase the inferred dispersions $\sigma_{\rm int}$ by $\textasciitilde 10$\%. 

This estimate is affected by the global velocity trend shown in Fig.~\ref{fig:gradV}. Here we estimate the velocity dispersion with the same method after subtracting the global velocity trend for the 15 tracers.
Since PNe have stellar progenitors, one would expect them to follow the stellar field dynamically. The GCs, while sharing the same potential, may be on very different orbits (e.g. have a different anisotropy and/or volume occupation). Indeed, the dispersion measured around the five trends plotted in Fig.~\ref{fig:gradV} ranges from 11.2 to 13.6~\kms, as opposed to the nominal value of $\sigma_{\mathrm{int}}=10.6^{+3.9}_{-2.3}$ km/s. Given the uncertainties associated with our limited set of 15 tracers, we find no significant evidence that their kinematics is the same as (or differs from) the trend observed in the stellar body. The velocity dispersion as estimated using the 15 tracers is lower than the velocity dispersion of the galaxy stellar body as reported in Sect.~\ref{sec:udgkinematics}, though consistent within uncertainties \citep[see also][]{Laporte2018}.

\subsection{DF2 dynamical mass}

We now turn to the estimate of the mass of DF2 using several estimators and either the velocity dispersion of the discrete tracers (GC and PNe, see previous section) or the stellar velocity dispersion of the galaxy (Sect.~\ref{sec:dispUDG}). The two dispersion values are consistent with each other within the derived uncertainties. As already emphasised, we do not necessarily expect them to exactly agree for at least two reasons : 1) these measurements do not cover the same physical (radial) range, with $\sigma_\mathrm{int}$ including tracers up to $\sim 80\arcsec$ or $\sim 3.5$~\re, while $\sigma_{\rm{DF2}\star}$(\re) was defined within one effective radius; 2) the small number of discrete measurements with relatively large uncertainties may lead to a significant under- (or over-) estimate plus
a systematic bias, as emphasised by \cite{Laporte2018}.

Recent mass estimators via the inferred effective radius \re\  and the associated velocity dispersion $\sigma_{\mathrm{los}}$ (within 1~\re) use relations proportional to $R_{\mathrm{e}} G^{-1} \sigma^2_{\mathrm{los}}$, with varying multiplicative constants.
A recent estimator was provided by \cite{Errani2018} with $M_{dyn} \left(< 1.8 \, R_{\mathrm{e}} \right) = 6.3 \cdot R_{\mathrm{e}} G^{-1} \langle \sigma^2_{\mathrm{los}} \rangle$ \citep[see also the earlier and similar estimator in][]{Amorisco2011}. This would lead to masses (within 1.8~\re, encompassing 87\% of the total modelled luminosity) of $3.6_{-1.4}^{+3.1} 10^8$~\Msun\ and $3.7_{-2.2}^{+2.5} 10^8$~\Msun\ when using $\sigma_{\mathrm{int}}$ and $\sigma_{\rm{DF2}\star}$(\re) respectively, and mass-to-light ratios M$_{dyn}$/L$_V$ of $3.7_{-1.5}^{+3.3}$ (resp. $3.9_{-2.6}^{+2.3}$) assuming Newtonian dynamics and a distance of 20~Mpc ($R_e \sim 2.2$~kpc). Following \cite{Courteau2014} (see their Sect.~5.2, and Table~2 of that section), we can also write an estimate of the mass within one effective radius as $M_{dyn} \left(R_{\mathrm{e}} \right) = 4 \cdot R_{\mathrm{e}} G^{-1} \sigma^2_{\mathrm{los}}$ \citep[see also][]{Wolf2010}, and this would lead to 10\% higher values for M$_{dyn}$/L$_V$. 
A closer distance of 13~Mpc ($R_e \sim 1.4$~kpc) would significantly increase the above-mentioned M$_{dyn}$/L by about 55\% (as inferred M$_{dyn}$/L values inversely scale with the assumed distance), with M$_{dyn}$/L$_V$ of 5.8 and 6.0 when using the GCs and PNe, or the galaxy stellar body.

A second way forward to estimate the dynamical mass of DF2 is to use Jeans modelling and predictions for the velocity dispersion of the galaxy within 1~\re. We thus apply the MGE formalism mentioned in Sect.~\ref{sec:udgkinematics} and solve the Jeans equations producing N-body realisations for a Sersic $n=0.6$ stellar body as in \cite{Emsellem2013}. Varying the anisotropy and inclination, we obtained modelled estimates for the observed stellar velocity dispersion $\sigma_{\rm{DF2}\star}$(\re) and can compare these with the observed MUSE value. Assuming a distance of 20~Mpc, and inclination angles between 90\degree (edge-on view) and 35\degree (close to face-on view), we derive corresponding M$_{dyn}$/L$_V$ between $\sim 3.5$ and 3.9 ($\pm 1.8$) to reproduce the 10.8~\kms stellar dispersion, consistent within 10\% with the ones provided above via direct mass estimators. This scales to M$_{dyn}$/L$_V$ from $\sim 5.4$ to 6.0 when using a distance of 13~Mpc.
A mass-to-light ratio of 4 within 1~\re\ corresponds to a mass of $\sim 2.2\, 10^8$\Msun\ within that radius.

Finally, it is worth noting that the new estimates of the velocity dispersions, both from the discrete tracers or the stellar body of the galaxy, are in broad agreement with the modified Newton dynamics (MOND) prediction, once the external field effect is properly taken into account \citep{Famaey2018}, as they predicted $\sigma_\mathrm{MOND} = 13.4^{+4.8}_{-3.7}$~\kms \citep[also cf.][]{kroupa2018}.  

\subsection{Origin of DF2}

The stellar velocity field of DF2 (see Sect.~\ref{sec:udgkinematics}) is clearly not consistent with a simple oblate axisymmetric system. The galactic kinematic axis is closer to the minor than the major photometric axis of DF2, making this a prolate-like rotation field. Such structures are not unheard of for low luminosity galaxies \citep[see e.g.][]{Ho2012, Kacharov2017}. 

The origin of such systems has been discussed over the last decade with help from numerical simulations. Several scenarios were put forward. Firstly, an accretion event or major merger may be an efficient way to produce prolate-like rotation \citep{Amorisco2014, Lokas2014, Fouquet2017}, something that was also called upon for massive galaxies \citep[see e.g.][]{Krajnovic2018, Graham2018, Ene2018} backed up by numerical simulations \citep[e.g.][]{Velliscig2015, Li2016, Ebrova2017, Li2018}. Secondly, in the case of DF2, another process could be tidal stirring by a more massive nearby galaxy.
Simulations by for example~\citet{Lokas2010} have shown that tides by a massive host can drive the density of a dwarf to triaxial configurations, and result in prolate-like kinematic signatures.
Simulations by \cite{Ogiya2018} suggest that tidal stripping on a radial orbit could explain the main morphological parameters of NGC1052-DF2 and a lowered amount of dark matter, even though one needs fine-tuned initial conditions to reach these results. Simulations by \cite{Penarrubia2008b} and \cite{Fattahi2018} show that tidal stripping can lower the internal velocity dispersions, but the dwarf galaxies still remain heavily dark-matter dominated. Moreover, tidal stripping may lead to out-of-equilibrium kinematics, increasing the observed velocity dispersion and leading to apparently high total mass-to-light ratios even in the absence of dark matter \citep[e.g.][]{Kroupa1997,KlessenKroupa1998}. Thus, tidal stripping may lead to very different scenarios, depending on the initial conditions and orbital evolution of the dwarf galaxy. More generally, prolate and triaxial systems arise naturally due to tides, as has been shown for example by \citet{Varri2009} by using self-consistent distribution functions in external tidal fields. This simply derives from the fact that Roche lobes are triaxial.

One hint that the galaxy may be perturbed and not phase-mixed is the detection by \cite{Trujillo2018} of an extended surface brightness feature towards the north in their very deep Gemini $g$-band image. We may further speculate that the discrepant velocities for PN3 (and possibly GC93) are associated with a dynamical disturbance.

\section{Summary and conclusions}
\label{sec:summary}

We have presented the kinematic study of the stellar body of NGC~1052-DF2, using new integral-field spectroscopy from MUSE, a first for this spectrograph and a galaxy at such low surface brightness. The large field of view of MUSE allows us to simultaneously study the global kinematics of the stellar body, and the systemic velocities of point sources such as globular clusters and nebular emitters. We illustrate the systematics, focusing primarily on a treatment of the dominating sky background in the MUSE datacube, but also by testing the impact of the spectral range as well as of using different spectral libraries.

We find that the receding velocity of DF2 is consistent with that of the mean globular cluster velocity. We reveal a significant velocity gradient throughout the MUSE field of view of about $\pm 6$~\kms within a radius of $\sim 20\arcsec$. The stellar velocity map is not consistent with a simple oblate rotator. The kinematic major axis is offset by about $70\degree$ from the photometric major axis, making this system a prolate-like rotator.

We measured spectra for seven globular clusters with sufficiently high S/N to be able to measure a reliable systemic velocity. Another compact source, with broadband colours consistent with a globular cluster, is confirmed to be a Galactic foreground star (likely a G-type star). We also detected the presence of three PNe within the DF2 field of view, and derived their radial velocities, two of them being in line with DF2 measured stellar kinematics.

We provide two velocity dispersion measurements, one from the set of clusters and one of the stellar body. First, we revise the velocity dispersion reported in the literature for this system, which was based on ten globular clusters. Folding in 1) two additional measured globular cluster velocities, 2) improved measurements for five others, 3) three PNe under the assumption that they are similar tracers of the potential, and 4) a precisely determined mean velocity of the whole system, we find a velocity dispersion of $\sigma_{\mathrm{int}} = 10.6^{+3.9}_{-2.3}$~\kms. 
Second, we measured the stellar velocity dispersion of the galaxy stellar body within one effective radius with $\sigma_{\star}(R_e) = 10.8_{-4.0}^{+3.2}$~\kms, with lower (resp. upper) limits of 5~\kms (resp. 12~\kms) when accounting for systematics associated with the covered spectral domains, and a full 95\% confidence interval of [0 - 21]~\kms.

While the dispersion estimates based on the stellar body and that of the discrete tracers are consistent with each other, an interpretation of these measurements should wait for a detailed modelling effort, which would include for example the prolate-like rotation suggested by our MUSE dataset. The present study generally demonstrates the power of integral-field spectrography and more specifically the capabilities of MUSE@VLT for low surface brightness systems. It should further motivate observational campaigns targeting dwarf galaxies in diverse environments, thus helping us to constrain their formation paths. In a companion paper \citep[Paper II,][]{Fensch2019}, we study the stellar populations of this low-luminosity extended system, compare it to the stellar populations of the globular clusters in this system, and discuss the origin of this galaxy. 

%-------------------------------------------------------------------
\begin{acknowledgements}
We thank Magda Arnaboldi, Souradeep Bhattacharya, and Glenn van de Ven for helpful discussions. RvdB thanks the participants of the Lorentz Centre Workshop ``The Bewildering Nature of Ultra-diffuse Galaxies'' for insightful discussions. We thank Lodovico Coccato for his extremely useful help on the use of \textsc{ZAP}. We thank Simon Conseil for useful correspondence on the use of \texttt{ZAP}, Laure Piqueras for her insight and support regarding mpdaf and both for general feedback on MUSE data reduction related topics, and we thank Benoit Epinat for his help with fitting the sky lines. We also thank Tim Oliver Husser, Philippe Prugniel, and Damien Le Borgne for their support pertaining to the stellar template libraries.
Analysing the MUSE data and writing the paper, we made use of Astropy (Astropy Collaboration et al. 2013), with heavy usage of the Python packages NumPy 
(Walt et al. 2011), iPython (Prez \& Granger 2007), SciPy (Jones et al. 2001), matplotlib (Hunter 2007), mpdaf \citep{Piqueras2017}, \texttt{emcee} \citep{emcee}  as well
as Spectres \citep{Carnall2017} and pPXF \citep{Cappellari2004, Cappellari2017}. We would thus like to thank all who designed and contributed to these excellent pieces of software, and most importantly made them available to the community.
\end{acknowledgements}

\bibliographystyle{aa} 
\bibliography{library} 

\begin{appendix}
\section{Spectral sampling and resolution}
\label{app:resprofiles}

In order to test the robustness of our results, we analysed two datacubes from the merging of the same MUSE pixel tables, only differing in their spectral sampling: one with a standard linear sampling of 1.25~\AA\ per pixel, and a second sampled following the logarithm of the wavelength with a scale corresponding to 30~\kms\ per pixel, taking advantage of the 28 dithered and rotated individual exposures. The latter removes the need to apply a second resampling when using pPXF (see Sect.~\ref{sec:vel}). Results obtained with these two datacubes are consistent with each other (within the quoted one-sigma uncertainties).

\begin{figure}
   \centering
   \includegraphics[angle=0,width=9cm]{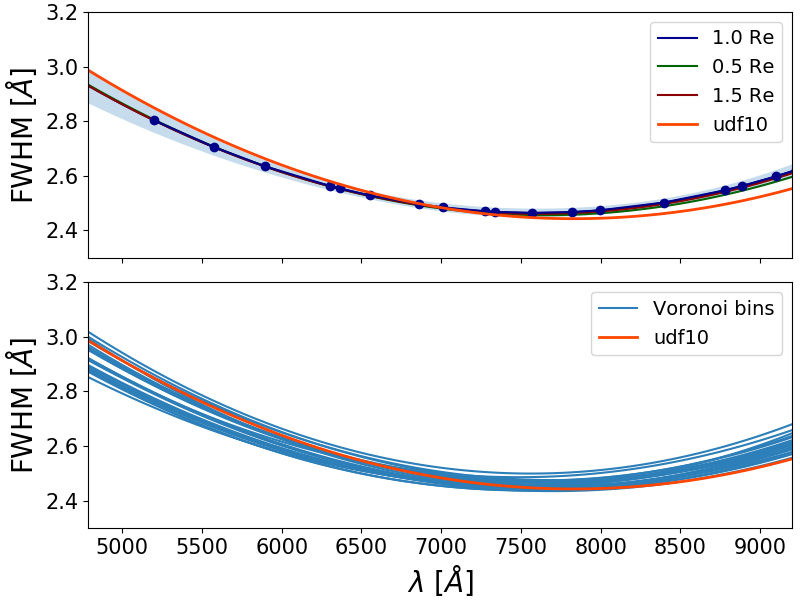}
      \caption{Spectral resolution profiles, given by the FWHM as a function of
      wavelength for the three apertures (top panel) and 21 Voronoi bins (bottom panel), as compared to a
      reference profile \citep[UDF10][]{guerou17}. In the top panel, the pale blue area illustrates the one-sigma
      uncertainty on the measured profile for the 1~\re\ aperture (the profiles for the three apertures being nearly
      indistinguishable).
              }
         \label{fig:app_resprof}
\end{figure}

As described in the main text (see Sect.~\ref{sec:measureveldisp}), the spectral resolution profile of the MUSE apertures and bins is key to the derivation of the stellar velocity dispersion. In Fig.~\ref{fig:app_resprof}, we show the variation over the different apertures (top panel) and Voronoi bins (bottom panel) of these profiles depending on wavelength, using a Gaussian approximation (parameterised by its FWHM) for the sky lines. All resolution profiles are overall very similar, but not identical, to the so-called UDF10 profile \citep{guerou17}, which is shown in Fig. A.1 as a reference.

\section{Robustness of sky subtraction method}
\label{app:skysubtraction}

For a correct interpretation of our results, it is essential that the signal of the galaxy is robustly separated from the sky signal. Especially in the low-surface brightness regime in which we are operating, this can be a challenge. Here are the three important parameters for the sky subtraction computation: 

\begin{itemize}
    \item {\bf Mask of the low-luminosity galaxy:} One needs to use a mask for which enough sky is available to sample well the variation of the LSF over the field of view and reduce the noise of the method, with the least possible contamination from the galaxy. We used a series of elliptical masks and chose the size corresponding to a convergence of the extracted flux of the galaxy and to the rise of noise (circularized radius: $30\arcsec$). This is the mask shown in the middle panel of Fig~\ref{fig:UDG}. In order to verify that the size of the mask does not affect the result, we also analysed cubes created with a larger masking of DF2, using a circularised radius of $36\arcsec$. 
    
    \item {\bf Number of eigenvectors:} by default, \textsc{ZAP} chooses the optimal number of eigenvectors to subtract to the datacube by looking at the inflection point of the variance curve. For our cube, \textsc{ZAP} computed that 80 eigenvectors are the optimal value. However, the higher the number of eigenvectors, the higher the probability to remove signal from the source. We studied the evolution of the extracted spectra with the number of eigenvectors and found a convergence starting at 30 eigenvectors. In order to verify the effect of this parameter on the results, we analysed cubes with three different numbers of eigenvectors: 30, 45, and 60.
    
    \item {\bf Window for the continuum subtraction:} before computing the singular value decomposition and the eigenvalues, \textsc{ZAP} filters out the continuum using a weighted median filter in a moving window whose size is chosen by the user. \citet{Soto2016} found that a width of 20-50 pixels is optimal. We extracted spectra using a range of values and found a convergence for widths between 30 and 50 pixels. In order to verify the effect of this parameter on the results, we analysed cubes with two different window sizes: 30 and 50 pixels.
\end{itemize}   

Thus we analysed 12 cubes, each with a different set of parameters for the \textsc{ZAP} sky reduction process. The resulting velocity measurements are all included in the systematic uncertainties quoted in the main tables of this work.

As a final test we also applied a method to subtract the sky background that is completely independent from ZAP, hereafter called ``conventional'' sky subtraction. For this we used a galaxy template that is constructed from the HST F814W image. We re-grided this image to the same pixel grid as the MUSE cube and convolved the data to the same PSF as the MUSE image quality. We fitted this template to each wavelength channel independently. In addition to this we fitted a constant (i.e. a plane) to represent the sky background emission in each channel. As expected, this method works well in regions that are devoid of bright sky emission lines. 

As a robustness test, we ran our kinematic analysis on the cube with the ``conventional'' sky subtraction. Whereas the spectra are cosmetically less clean than those extracted from the ZAPped cubes, the differences occur mostly near bright sky lines; in these regimes, ZAP performs remarkably better. However, those spectral regions are masked when performing our kinematic measurements, and we find that none of the derived velocities are significantly changed compared to extractions from the ZAPped cubes. The systematic differences are included in the second errors that are quoted in Table~\ref{table:sysvels}.

\section{Kinematics of NGC1052-DF2}
\label{sect:vbins}

We provide here (Table~\ref{table:vbins}) the list of Voronoi bins used to map the kinematics of NGC1052-DF2, and the associated velocities, including values both from pPXF and cross-correlation methods. 

\begin{table}
\caption{Positions and systemic velocity measurements of DF2 in Voronoi bins 
(using the eMiles library). There are 21 bins, for which we provide the relative positions of the nodes and corresponding velocities as measured with the pPXF fitting and cross-correlation methods. The X- and Y coordinates refer to angular offsets with respect to the centre, RA=40.44542$^{\circ}$, Dec=$-$8.40333$^{\circ}$, in the western and northern directions respectively. In columns 5 and 6 (minV, maxV) we also provide systematics when using various setups for the sky subtraction (see Appendix~\ref{app:skysubtraction}). }   
\label{table:vbins}
\centering          
\begin{tabular}{r r r r r }     % 4 columns 
\hline\hline       
Bin & X node & Y node & V (pPXF) & V (CC) \\
& [arcsec] & [arcsec] & [km/s] & [km/s] \\
\hline   
 1 &  -0.10 &  -1.09 & $1793.5_{-2.6\,-1.1}^{+2.5\,+0.8}$  & $1795.4_{-2.1\,-2.8}^{+5.2\,+2.6}$ \\
 2 &  -4.09 &   2.01 & $1793.8_{-3.5\,-0.5}^{+3.0\,+1.1}$  & $1797.5_{-3.0\,-3.3}^{+3.0\,+2.6}$ \\
 3 &  -2.60 &   6.97 & $1793.1_{-3.0\,-0.8}^{+2.9\,+1.1}$  & $1794.9_{-5.4\,-2.2}^{+0.7\,+2.1}$ \\
 4 &  -8.59 &   7.65 & $1790.1_{-4.1\,-2.9}^{+4.7\,+0.5}$  & $1793.2_{-4.8\,-5.4}^{+6.4\,+2.8}$ \\
 5 &  -9.22 &   2.36 & $1790.3_{-3.0\,-1.9}^{+2.8\,+0.7}$  & $1790.5_{-2.9\,-5.4}^{+2.5\,+1.4}$ \\
 6 &  -4.77 &  -3.42 & $1795.1_{-2.7\,-0.8}^{+2.5\,+0.5}$  & $1795.7_{-2.2\,-3.6}^{+4.1\,+1.8}$ \\
 7 &   2.48 &  -6.35 & $1796.0_{-2.7\,-0.5}^{+4.0\,+0.1}$  & $1796.7_{-4.3\,-1.5}^{+2.6\,+1.2}$ \\
 8 & -11.89 &  -1.79 & $1790.5_{-3.0\,-1.2}^{+3.1\,+1.0}$  & $1791.3_{-4.0\,-5.2}^{+3.5\,+2.2}$ \\
 9 & -19.05 &  10.00 & $1789.6_{-5.0\,-3.5}^{+5.4\,+1.1}$  & $1793.2_{-1.4\,-9.0}^{+2.5\,+4.7}$ \\
10 & -21.88 &  -7.17 & $1799.1_{-3.4\,-2.8}^{+3.2\,+0.4}$  & $1801.7_{-3.5\,-3.7}^{+3.7\,+0.8}$ \\
11 &  -5.98 &  -8.40 & $1797.3_{-4.3\,-2.1}^{+2.8\,+0.0}$  & $1798.2_{-3.3\,-3.7}^{+2.5\,+1.8}$ \\
12 &   1.16 &   2.95 & $1789.6_{-2.9\,-1.0}^{+2.3\,+0.5}$  & $1790.2_{-1.8\,-2.2}^{+4.4\,+1.0}$ \\
13 &  -7.13 &  21.40 & $1788.5_{-3.9\,-2.6}^{+4.4\,+1.2}$  & $1792.1_{-2.3\,-7.6}^{+2.9\,+4.1}$ \\
14 &   4.00 &  12.60 & $1790.6_{-2.5\,-1.2}^{+3.1\,+0.7}$  & $1792.1_{-2.3\,-5.2}^{+2.9\,+2.1}$ \\
15 &   5.67 &   5.11 & $1793.2_{-2.9\,-0.1}^{+2.6\,+0.4}$  & $1793.6_{-1.9\,-1.0}^{+2.8\,+0.8}$ \\
16 &   1.24 & -12.40 & $1795.1_{-3.3\,-0.1}^{+4.0\,+0.4}$  & $1798.6_{-3.0\,-1.1}^{+1.8\,+1.8}$ \\
17 &   7.29 &  -2.28 & $1793.9_{-2.1\,-0.0}^{+2.0\,+0.7}$  & $1794.5_{-4.1\,-1.8}^{+0.4\,+1.8}$ \\
18 &   8.61 &  23.13 & $1787.6_{-3.3\,-1.1}^{+3.2\,+2.2}$  & $1789.1_{-4.3\,-3.0}^{+4.4\,+2.9}$ \\
19 &  12.55 &   9.36 & $1790.8_{-3.0\,-0.9}^{+2.7\,+0.1}$  & $1789.4_{-1.5\,-1.2}^{+3.3\,+1.7}$ \\
20 &  12.77 &  -0.15 & $1797.1_{-2.9\,-1.1}^{+2.6\,+0.0}$  & $1798.2_{-5.4\,-2.1}^{+2.6\,+0.4}$ \\
21 &  11.18 &  -9.71 & $1803.1_{-3.2\,-1.1}^{+2.0\,+0.7}$  & $1803.6_{-2.6\,-1.8}^{+4.1\,+5.4}$ \\

\hline                  
\end{tabular}
\end{table}

\end{appendix}
\end{document}